# ARTICLE

# Bringing ultimate depth to scanning tunnelling microscopy: deep subsurface vision of buried nano-objects in metals

Oleg Kurnosikov*[a, b], Muriel Sicot[a], Emilie Gaudry[a], Danielle Pierre[a], Yuan Lu[a], Stéphane Mangin[a]

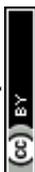

**Abstract**

A method for subsurface visualization and characterization of hidden subsurface nano-structures based on Scanning Tuneling Microscopy/Spectroscopy (STM/STS) has been developed. The nano-objects buried under a metal surface up to several tens of nanometers can be visualized through the metal surface and characterized with STM without destroying the sample. This non-destructive method exploits quantum well (QW) states formed by partial electron confinement between the surface and buried nano-objects. The specificity of STM allows for nano-objects to be singled out and easily accessed. Their burial depth can be determined by analysing the oscillatory behaviour of the electron density at the surface of the sample, while the spatial distribution of electron density can give additional information about their size and shape. The proof of concept was demonstrated with different materials such as Cu, Fe, W in which the nanoclusters of Ar, H, Fe and Co were buried. For each material, the maximal depth of subsurface visualisation is determined by the material parameters and ranges from several nanometers to several tens of nanometers. To demonstrate the ultimate depth of subsurface STM-vision as the principal limit of our approach, the system of Ar nanoclusters embedded into a single-crystalline Cu(110) matrix has been chosen since it represents the best combination of the mean free path, smooth interface and inner electron focusing. With this system we experimentally demonstrated that Ar nanoclusters of several nanometers large buried as deep as 80 nm can still be detected, characterized and imaged. The ultime depth of this ability is estimated as 110 nm. This approach using QW states paves the way for an enhanced 3D characterization of nanostructures hidden well below a metallic surface.

**New concept**

It is a common belief that the target of a scanning tunnelling microscopy and spectroscopy (STM/STS) analysis is exclusively a surface or objects on it. With our new concept, we show that the STM/STS method can also be applied to characterize hidden nano-objects and nanostructures buried up to 100 nm below a surface. Although using the conventional STM/STS approach and by analysing subtle surface deformations or slight deviations in the local electronic state density the subsurface objects could be recognized in some specific cases, their depth is usually limited to a few atomic layers. In contrast, our approach uses quantum well (QW) states formed by the confinement of delocalized electrons between a surface and buried nano-objects. The use of QW increases the ultimate depth of detection by one order of magnitude, which is limited by the mean free path of delocalized electrons. Spatial variations of QW states and oscillatory STS spectra allow the location, depth, and size of the buried nanoclusters to be determined, as well as indicate their shape. The new concept paves the way for non-destructive STM characterization of buried nanostructures and triggers the development of 3D subsurface nano-analysis with STM.

## Introduction

Scanning tunnelling microscopy (STM) and spectroscopy (STS) have become a widely used technique for imaging surface structures with atomic resolution [1] and determining the local density of electronic states (LDOS) of surface atoms, molecules or nanostructures [2]. The extremely high surface sensitivity of this technique is a direct consequence of a strong exponential decay of the probability of tunnelling selecting the atoms closest to the tip. Therefore, it is commonly believed that an STM can be used exclusively for characterization of a surface as well as atomic, molecular and nanostructures on it. For profiling structures below a surface, the STM still can be used together with sample-destructive post-processing like milling [3] or cleaving the sample [4]. This destructive post-procession forms a new surface from deeper layers of materials or in cross-sections and then STM performs the surface characterization on

a. Université de Lorraine, Institute Jean Lamour, France.
b. ex. Technische Universiteit Eindhoven, department of Applied Physics, Netherlands.
* Contact address: Oleg.Kurnosikov@univ-lorraine.fr

**ARTICLE**  Nanoscale Horizons

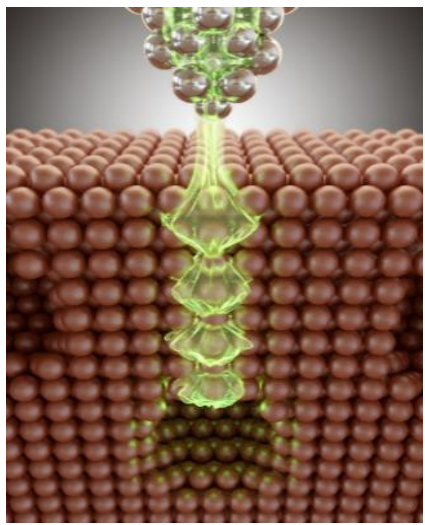

Fig.1 Illustration of the concept of subsurface STM vision based on the near-surface resonances of non-localized electronic states of the matrix material.

them. This way for accessing deeper structures is widely used also for depth profiling study with transmission electron microscopy [5], scanning electron microscopy [6], as well with electron spectroscopy [7]. However, this sample-destructive technique has drawbacks as the interfaces at embedded structures can be damaged during the sample post-processing. Frequently, the destructive post-processing has to be performed in several stages or with a set of the samples for detailed characterization.

In this article, we develop the subsurface STM/STS vision through a metallic surface for characterisation of the nano-objects buried up to 100 nm below the surface excluding any sample-destructive post-processing like milling or cleaving. Our approach allows to see nano-objects at different depths with one STM scan and to determine their locations and depths for a wide variety of conductive materials of substrates. In addition, the size of the subsurface nanostructures can be determined or estimated if the electronic structure of the substrate is suitable and known. In this case, the spatial distribution of scanned anomalies can even provide more insight into the shape of buried nanostructures. Avoiding any sample-destructive post-processing gives an opportunity to follow the evolution of the systems with embedded nano-objects in time under different activation processes.

The ability of the STM to indicate in its images the presence of subsurface objects was demonstrated previously, but it was not widely used for subsurface characterisation. There are some examples of STM images showing that tiny deformations of the surface or an appearance of additional electronic states at the surface can be attributed to subsurface atoms or nanostructures buried [8, 9, 10] one or a couple atomic layers below a surface. In semiconductors, near-surface impurities are recognized as extra features superimposed onto the STM images of a surface [4, 11-14]. Such effects can become very spectacular in the case of single dopants in semiconductors since the embedded charges and states of the dopants are not screened over the distance of several nanometers and therefore contribute directly to the LDOS at the surface [15]. As a contrary, in materials with a very high electron density like metals, such subsurface impurities and defects buried several nanometers underneath can not be detected as easily since their states and charges are screened already over the interatomic distances. As a consequence, subsurface STM vision at the depth on the nanometre scale in metals is considered to be unrealistic.

In this article, we oppose this statement by presenting an experimental evidence of the extremely deep subsurface vision in metals with STM. Instead of direct detection of the localized electronic states belonging to the buried impurities or nanostructures we use bulk non-localized states of the host material (Fig. 1). The electron density of the bulk between the surface and the buried nano-object gets the oscillating component and this component can be probed with STS at the surface near the location of the hidden nanoobject. Because the length of delocalization of these states can be very large, a deep subsurface structure can be detected.

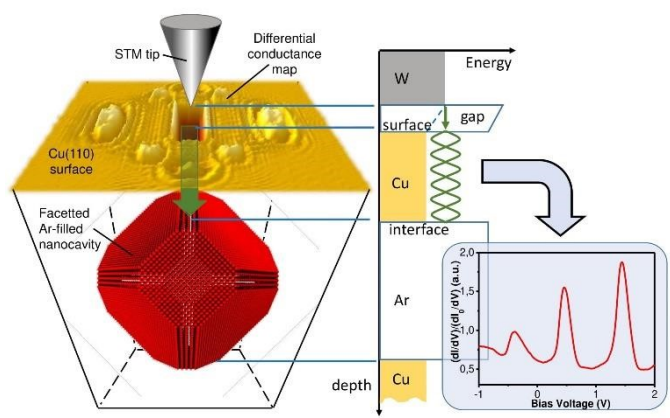

Fig. 2. Schematic drawing of realisation of the subsurface detection of hidden nanocluster. Left panel illustrates the configuration of the system and lateral distribution of the surface conductance induced by the subsurface nanocluster with a facetted interface. The middle panel is the corresponding energy diagram leading to the measured oscillatory behaviour of the normalized differential conductance in STS spectrum displayed in the bottom right panel.

Previously, several STM experiments reported the use of LDOS oscillations in a metallic layer for the subsurface detection and thus confirm the validity of this approach. For example, the steps at Si substrate covered with a flat Pb layer were imaged by analysing LDOS variation on the flat surface of lead [16-23] whereas the same effect was also observed with thin films of Cd [24]. A similar concept has been used by Weissman *et al.* [25] and Kotzott *et al.* [26]. They were able to locate subsurface impurity atoms buried several atomic layers below the Cu surface by treating the diameter of oscillating LDOS rings. The spatial oscillations of LDOS induced by electron scattering at a subsurface atom is theoretically described by Avotina *et al.* [27] and Lunis *et al.* [28]. There are also a few examples of imaging the subsurface nanoobjects showing stronger signal due to efficient reflection. The systems are represented by nanocavities formed in metals like Cu, Pb, Ag or Al and filled in by noble gases [29-36]. However, the typical depth of location of subsurface structures in these reports does not exceed a couple of nanometers while the signals still reveal remarkable and measurable oscillations in LDOS. This indicates that the limit in the depth has not been achieved.

In this article, we show that the limit of subsurface vision can be pushed in one or two orders more than previously reported. For illustration, we exploited an ideal system which was used in pioneering works of Adam *et al.* and Kurnosikov *et al.* [33-35], however they reported only some particular results. Understanding of all limiting factors allowed us to achieve the depth of subsurface vision in our experiments up to 80 nm and estimate the ultimate depth of detection of around 110 nm.

## Principles

As well known, the LDOS could be accessed by means of STS at the surface [2]. The LDOS at the surface is usually determined by the band structure of material at corresponding crystallographic orientation, by the states associated with the crystal termination, so-called surface states, and by localized states of single atoms or molecules in- and on the surface. The last type of states is mostly used for the visualization of atomic structure in STM images, whereas the two first ones usually contribute to a homogeneous background, which meets much less of interest in STM applications. However, in some cases the delocalized states can reveal interesting features in the STM/STS images. Regarding a system with *surface delocalized* states, for example the stepped surface of Cu(111) or the same surface with surface impurities, standing wave patterns of electron density can be observed on STS maps [37]. The standing waves surround the impurity atoms or group near the step edges since they originate from the scattering of delocalized surface electrons on these objects and electron wave interference. Exactly the same mechanism of formation of standing waves of electron density should be realized also in volume for *bulk delocalized* electronic states in case of the scattering or reflection from the objects located in the bulk, *i.e.* below the surface. In this case, however, the deep subsurface object cannot be reached with the STM probe and therefore does not appear in STM images. Nevertheless, the perturbation of the spectra of delocalized bulk states due to subsurface scattering can be observed *at the surface* as well. The perturbation of the spectra appears as superimposed electron density oscillations across the surface as well as the oscillation of electron density probed in a selected location versus energy since the electrons at the scattering obey a wave equation. Considering the scattering from a bigger subsurface object, like an interface with some buried nanocluster and a free flat surface, one can expect a multiple electron reflections between them. In this case, the interface and surface serve as boundaries partially confining the delocalised bulk electrons. The partial confinement could lead to a formation of localized near-surface quantum well (QW) resulting in much stronger quasi-periodical variation of the electron density. A schematic representation of such a geometry and its corresponding energy landscape is given in Fig.2. Therefore, the observation of *oscillatory component* in LDOS revealed by STS above the buried nanostructure can be used for the detection of buried nanoparticle. Moreover, by analysing the period of oscillation $\Delta V$, it would also be possible to determine the depth *d* of confinement of the bulk states that is also the depth of the corresponding facet of the subsurface structure, using the following formula:

$$d = \pi (dE/dk) \cdot 1/(e\Delta V) \quad (1)$$

where d$E$/d$k$ is the derivative of energy $E$ by the wave vector of electrons $k$ in the appropriate direction deduced from the band structure and $e$ is the electron charge. Beside the depth of location, also the shape or size of the subsurface structure can be reconstructed considering the lateral variation of LDOS with a corresponding model [33-34].

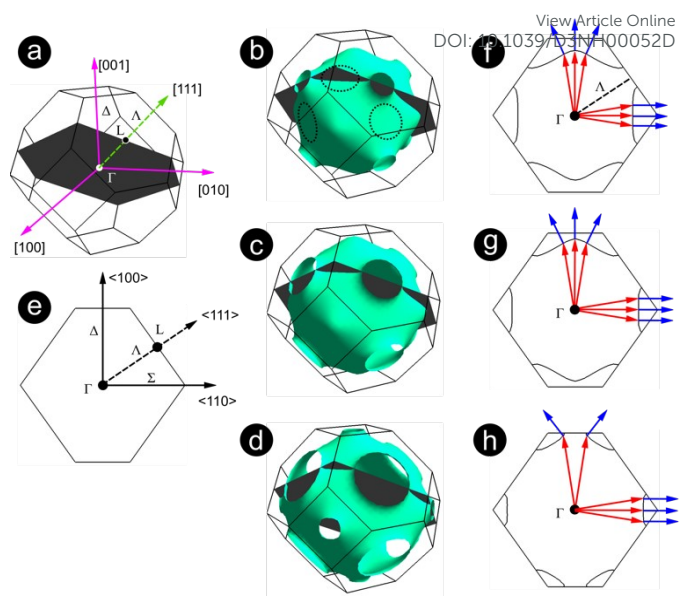

Fig.3 Principles of electron focusing in Cu: (a) First Brillouin zone (BZ) of an *fcc* lattice with relevant high-symmetry *k*-points, directions and lattice vectors. The grey plane is [1/2, 1/2, 1]; (b-d) Iso-energetic surfaces obtained from the calculated band structure of Cu in reciprocal space at (b) Fermi level, $E_F$ (Fermi surface), (c) $E_F$ + 1 eV and (d) $E_F$ +2 eV. The encircled areas in (b) highlight the saddle zones close to the <110> directions that are locally almost flat. (e) Cross-section between the grey plane and the first BZ in (a); (f- h) Cross-sections related to the grey plane [1/2, 1/2, 1] in (a, e) at the corresponding energies (f) $E_F$, (g) $E_F$+1 eV, and (h) $E_F$+2 eV, respectively. Red arrows illustrate *k*-vectors in the <110>, <100>, and close to them directions. Blue arrows indicate the directions of the group velocity. The propagation of electron waves with *k*-vectors around the <110> are almost parallel, whereas it diverges for electron packages with *k*-vectors close to the <100>.

Thus, the ability to detect and characterize the subsurface nanoclusters is based on the presence of the oscillatory component of LDOS in the STS signal. The amplitude of this component and its oscillation period decrease when $d$ increases. Therefore, an ultimate depth of detection would be defined as the depth when the oscillations are no longer recognizable in STS measurements.

Several factors are responsible for the formation of oscillations and thus determining the ultimate depth of STM vision, namely the electronic coherence length, interface roughness, and the electronic band structure of the host material, as discussed below.

First, since the subsurface vision is based on the electron interferences, the oscillation whose period is used to determine the depth $d$ as expressed in formula 1 can occur if the depth does not exceed the coherence length of the corresponding states. This coherence length is known to vary with the electronic mean free path (MFP). This last parameter, which is material-dependent, is limited by electron scattering on phonons, impurities or due to other collective interactions. Therefore, it varies with temperature T, impurities concentration, and electron energy. Overall, the MFP ranges from a few nanometers to several tenth of nanometers [38, 39]. Temperature modifies MFP such that MFP gets even much longer with temperature decrease. It is also noteworthy that temperature affects the broadening of the LDOS oscillation peaks belonging to each QW resonances. As a consequence, temperature will have an influence on the ultimate depth detection.

Second, the coherence can also be partially or completely lost at the reflection due to a rough interface. Therefore, buried nanostructures forming atomically flat interfaces with the host material are more likely to be detected with STM.

Third, the electronic band structure of the host material is important as well in determining the ultimate depth of detection. Its angular distribution is responsible for the divergence and decay of the electronic state packages since an STM probes the states with some variety of $k$-vectors in the three-dimensional reciprocal space. The decay depends on the direction of electron wave packages propagation in the bulk. As well known, the direction of propagation of wave package in the bulk at low energy is determined by the vector of group velocity $\mathbf{v_g(k)}$:

$$\mathbf{v_g(k)} = \nabla_k E(\mathbf{k}) \quad (2)$$

where the electronic dispersion relations $E(\mathbf{k})$ can be determined theoretically or experimentally in the form of electronic band structures [40]. According to formula (2), at a specific energy $E_0$ that can be experimentally set with STM choosing a bias voltage $V_0$ such that $E_0=eV_0$, in reciprocal space, the group velocity $\mathbf{v_g}$ is perpendicular to the iso-energetic surface of the host material. In case the bias is set to zero, the energy dispersion surface corresponds to the Fermi surface. In previous works [27, 28, 41, 42], it has been shown that the contours of the Fermi surface affect the direction of the group velocity. Specifically, when injecting electrons into directions where the Fermi surface is almost flat, the group velocity stays parallel for some close range of $\mathbf{k}$. This effect has been called the focusing effect. On the contrary, where the Fermi surface has a more spherical shape, the group velocity tends to diverge and as a consequence, the propagation directions are more spread. This has been illustrated experimentally in the case of single Co atoms embedded into a Cu(111) single crystal and theoretically explained using the Fermi surface of Cu [25, 28]. However, as the period of LDOS oscillations versus bias voltage is needed to determine the depth at which the object is buried, it is of paramount importance to not only consider the Fermi surface but contours at all energies within the range of the measure typically corresponding up to 2eV for a regular STS spectrum. Therefore, we have calculated iso-energy surfaces for Cu in the First Brillouin Zone shown in Fig.3(a) setting the energy to the Fermi level $E_F$, $E_F$+1eV, $E_F$+2eV as displayed in Fig. 3(b-d), respectively. In good agreement, with previous studies [25, 27, 28], the Fermi surface of Cu bear very flat regions with strongly suppressed curvature in <110> directions whereas necks resulting from band gap opening are observed in <111> directions around L points. In addition, energy bumps can be found in <100> directions around X points. Those three features remain valid within the range of 0-2eV as displayed in Fig. 3(b-d), respectively though the flat areas around Σ shrink with increasing energy due to the larger opening of the band gaps at L points. In order to better understand the influence of such contours on the focusing effect, we now discuss the anisotropy of the propagation of electronic waves in the STM experiment by looking at the group velocity as a function of the energy. To do so, we have chosen to consider a cross section between the plane [½ ½ 1] which contains the main 3 high-symmetry directions (see grey plane in Fig. 3(a) and its planar presentation in Fig. 3(e)) and the constant energy surfaces of Fig.3(b-d). The resulting cross-sections are displayed in Fig. 3(f-h) where red arrows illustrate $\mathbf{k}$-vectors whereas blue arrows indicate the group velocity giving the indication of the propagation direction, which is perpendicular to the contour displayed as black solid line. Obviously, upon increasing energy (i.e. bias voltage in STM experiment), the group velocity $\mathbf{v_g}$ remains parallel to the <110> direction for the $\mathbf{k}$-vectors grouping around <110>. However, if the corresponding $\mathbf{k}$-vectors lie around the <100> direction, the divergence of $\mathbf{v_g}$ occurs due to the presence of an energy bump in the reciprocal space. This translates into the fact that if electrons are injected into a (110)-cut surface, they will propagate mostly following that direction keeping the same amplitude over a long distance. However,

**Nanoscale Horizons**

**ARTICLE**

their injection into the (001)-cut surface will result in a stronger decay due to the spatial divergence of corresponding propagating states. This discussion could be generalized to any other materials: flatness of the contour of iso-energetic surface would lead to a focusing effect of the injected electrons [41, 42].

Since the local flatness of the iso-energetic surface of copper in <110> allows to concentrate the electronic states in this direction, the oscillating STS signal should be more pronounced and it would be worth to use this direction for enhancing the ultimate depth of the subsurface electron vision.

## Experiment

To demonstrate the subsurface STM-vision as well to illustrate its ultimate depth, we used several matrix materials such as Fe, W, and Cu with two different types of embedded nanoclusters, namely the insulating clusters of Ar and H, and the conductive nanoclusters of Co and Fe. Moreover, the systems of Fe/MgO and Bi/Fe were used to confirm that the same approach can be used in characterization of buried interfaces (Supplementary materials). The non-conductive nanoclusters represent the ideal case of total reflection at the interface with the matrix material. Since the atomic electronic structure of Ar and H does not play any remarkable role in the electronic processes we consider, we can reasonably treat these nanoclusters in metals also as nanocavities with Ar or H filling. On the contrary, Fe and Co nanoclusters buried in Cu provide only partial electron reflection since the nanoclusters are conductive. However, the mismatch of k-vectors in Cu and Co or Fe yields a reflection sufficient to realise the subsurface vision with this kind of non-ideal system. The band structure of all these materials is well-determined and the sample preparation is straightforward.

The Ar nanoclusters in copper and iron are formed by irradiating the Cu(110), Cu(001) and Fe(001) surfaces with a dose of $2 \cdot 10^{17}$ cm$^{-2}$ of Ar$^+$ ions with a kinetic energy of 5 keV and 1 keV, respectively. This ion energy was specifically chosen since it leads to a penetration of Ar into Cu up to 10 nm and a little bit less in Fe. In order to recover an atomically flat surface of Cu after irradiation with Ar, a moderate annealing at 1050K for 5 min was performed whereas Fe was annealed at 700K for 20 min. During this post-annealing, Ar segregation occurs. As a consequence, a significant part of the implanted Ar atoms reaches the surface and leaves the sample. However, some amount of the embedded Ar remains in the substrate forming nanoclusters of a typical size of around 2 - 10 nm. The process of formation of the nanoclusters has been proved by XPS analysis [34, 36]. Additionally, the XPS spectra at the nanoclusters formation can be found in supplementary materials. Formation of H nanoclusters is performed in a similar way by irradiating the W(111) surface with H$^+$ ions with a kinetic energy of 0.7 keV. A post-annealing at c.a. 2000K has been performed in a flash mode. To form Co and Fe nanoclusters embedded in Cu, half of a monolayer of Co or Fe was deposited on clean Cu(001), Cu(111) or Cu(110) single-crystalline surfaces. The annealing at 550K led to the formation of small nanoislands embedded into the first two layers of copper [9]. Then, they were buried further by extra deposition of Cu on top. Similarly, as we needed Ar nanoclusters buried deeper than those obtained after the implantation and the first thermal treatment, we deposited extra layers of Cu with increments of 10 nm thickness. A slight annealing to recover a flat Cu surface followed each post-deposition. The deposition was done with a calibrated effusion cell EFM3. Between each step, the sample was scanned with STM to visualise the nanoclusters.

The STM/STS measurements were done using a low-temperature STM microscope "Omicron" at temperatures of 77K and 4.7K under ultra-high vacuum. To get reliable STS measurements, we carefully conditioned the STM tips by means of local electron beam annealing. This process was controlled by checking the cold field emission from the tip. With this procedure, we were able to obtain repeatable and reliable oscillatory behaviour of the LDOS. Nevertheless, we cannot exclude some dispersion among STS spectra that could originate from the variety of the tip states at the very apex of different STM probes.

## Results and discussion

Since the van der Waals binding of Ar and H atoms in the clusters is much weaker than the binding of Cu, W, or Fe atoms in the metallic crystalline lattice, the shape of the subsurface nanoclusters is determined to a great extent by the anisotropy of the surface tension of the crystal. This should provide a Wulff construction [43] of the nanoclusters in the inversed form: a hollow polyhedron with the determined facets shape. Figure 2 presents the expected shape of the last atomic layer of copper at the interface with the Ar nanocluster deduced from an estimation of the anisotropy of the Cu surface energy as well as from some experimental data [44, 45]. The atomically flat facets at the interface serve as ideal reflectors for the delocalised electron states outside the nanocluster. The nanoclusters formed in Cu crystal with the (001) and (110) cut have the same shape, however their relative orientation to the surface is different: the upper facets confining the electrons are different, therefore the electronic states used for QW formation also are different. Similarly to fcc Cu sample, the faceted nanoclusters of Ar or H are formed in bcc Fe and bcc W crystals as well. Yet, their particular shapes differ from each other since all crystals have different surface energies for the principal facets.

However, the shape of Co and Fe nanoclusters embedded in Cu does not follow the Wulff constriction since the relatively low annealing temperature does not provide a local quasi-equilibrium state at the sample formation. Therefore, the

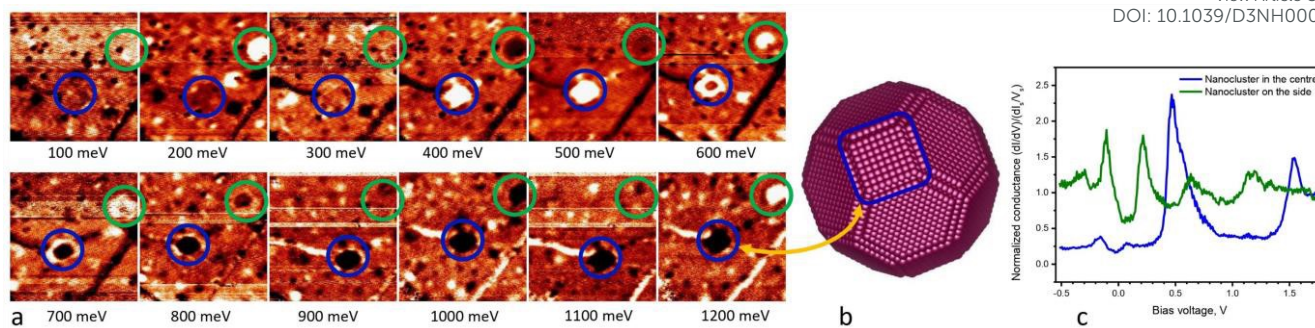

Fig. 4 (a) - STS maps as a function of bias voltage recorded on the same area of a Fe(001) sample with embedded Ar nanoclusters. Size: 20 x 20 nm$^2$. A periodical variation of contrast with bias in two large spots is highlighted by a blue and a green circles. The spot shape presents some rectangular-like features; (b) - Expected shape of a buried nanocluster represented by the first interface layer of Fe with the square upper (001) facet; (c) – STS normalized conductance curves showing the resonance peaks or oscillations in the two spots encircled in (a).

clusters keep their flat pancake-like shape obtained at first deposition on the surface and slight annealing. The roughness of the interface could however be affected.

### 1. Fe(001) with embedded Ar nanoclusters

After preparation of the Fe(001) sample with buried Ar nanoclusters, large atomically flat terraces are observed while some steps, single defects and point-like contaminations were always presented in the STM images. STS mapping (Fig. 4) reveals these defects as black and somewhere white points of deviating conductance. However, in some locations, the surface conductance maps show also large spots which cannot be associated with point surface additives or imperfections well recognisable also on standard STM images. The contrast in these locations appears to be supressed but also enhanced or unchanged at the selected bias voltage. The contrast of spots varies periodically with the bias voltage. However, the period is specific for each spot. This is usually not the case for the objects like the surface contamination observed. Figure 4 (a) shows the series of STS maps with two of such spots (encircled) periodically changing their contrast with bias voltage. Remarkably, the spot shape has some square features. The STS plots (Fig. 4 (c)) measured in the centre of the encircled areas indeed shows the quasi-periodical variation of the LDOS. The blue plot measured on the central spot (encircled in blue in Fig. 4 (a)) shows resonances at -0.15 V, 0.5 V and 1.5V; while the green plot corresponding to the second spot (encircled in green) reveals peaks at -0.3, -0.1 V, 0.2 V, 0.65 V, and 1.2 V. All these peaks manifest the QW resonances due to electron confinement between the surface and a local interface, namely one of the facets of buried Ar nanocluster. The shape of the nanocluster is shown in Fig. 4 (b) and was calculated using the surface energy of Fe according to Wulff's approach. The shape of the spot obviously reflects the square-like (001) facet. The depth of location of the upper facet of the nanocluster is derived from formula 1 using the corresponding band structure of Fe [46]. The upper facet of the buried nanocluster in the centre of Figure 4 (a) (encircled in blue) is located at a depth of 1.3 nm below the surface, whereas the corresponding facet of the second nanocluster (encircled in green) is located at a depth of 2.5 nm. The distance between the resonance peaks of the shallow nanocluster is larger and the peaks themselves show higher intensity. The asymmetric and narrow peak shape indicates the multiple reflection of the confined electrons, the condition at which the QW states are formed.

### 2. W(111) with embedded H nanoclusters

STS maps of the W(111) surface also show similar characteristic spots with periodically varying contrast with bias voltage, despite the presence of surface contamination. In the case of the (111) cut and contrary to the previous example, the spots reveal a triangular shape surrounded by additional three ray-

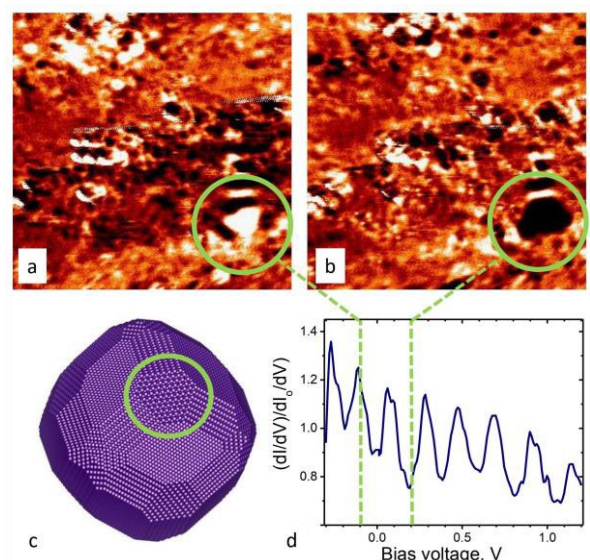

Fig. 5 (a, b) –STS maps recorded on the same area of the W(111) surface contained embedded H nanoclusters at a bias voltage of (a) -0.1V and (b) 0.2 V. Size: 20 x 20 nm$^2$. The triangular feature showing an opposite contrast is encircled; (c) - Expected shape of an H nanocluster represented by the first interface layer of W. The truncated triangular (111) facet is encircled; (d) – Normalized conductance versus bias voltage measured in the centre of the spot encircled in (a, b). The bias voltages corresponding to opposite contrast in (a) and (b) as indicated by dashed lines can be attributed to a maximum and a minimum of the oscillating conductance plot.

like features which contrast varies spatially. Figure 5 (a, b) shows STS maps of one same area measured at two bias voltages. Similar to the Fe(001) sample, these features can be attributed to the upper (111) truncated triangular facet of the nanocluster interface as depicted and encircled in Fig. 5 (b).

The STS curve measured in the centre of the spot encircled in Fig. 5 (a) reveals oscillations over a wide range of bias voltages. These oscillations can be associated with QW states. However, since the shape of the peaks becomes more symmetrical and the relative amplitude is lower, in comparison with Fe sample (Fig. 4 (c)), we can infer a partial loss of coherence of the confined electronic states and only a few reflections in the confining region. Nevertheless, a high contract on the STM maps allows to determine the exact lateral location of the buried nanocluster, despite the masking image of surface contamination, while the well-resolved minima and maxima in the oscillatory curves determine the depth of reflecting nanostructure. The depth of the facet of buried nanocluster of 5.2 nm is derived from the maxima positions and the band structure of tungsten [47] using equation 1.

These two examples confirm the generalization of our approach in application to various materials. These examples also show that the amplitude of the oscillating STS signal decreases with the depth. If we proceed with deeper and deeper nano-objects, a specific depth would be reached at which the oscillations of LDOS in QW would no longer be detectable. Therefore, below we address the question about the ultimate depth of STM detection.

## 3. Cu(110) with embedded Ar nanoclusters: ultimate depth

As discussed above, the QW formation and capability of subsurface STM vision are determined by the material characteristics such as the MFP, the electron coherence, the surface and interface quality, and the electronic states of the materials used as the matrix and for nano-objects. Considering those factors, the best candidate to demonstrate the ultimate depth of the subsurface STM vision would be Ar nanoclusters embedded in Cu(110). Besides a rather long MFP in copper which increases at low temperature, and a smooth interface with nanocluster, we also can benefit from the electron focusing effect discussed in the introduction. On the other hand, the (110) facet is the smallest one providing less efficient electron confinement between the interface and the surface. However, since focusing in the <110> direction seems to be a countable factor, it might compensate this disadvantage thus leading to an intense oscillating reflected signal. Therefore, with this type of sample we would like to reach experimentally the maximal practical depth and subsequently estimate the ultimate depth. Such an experiment would indicate the principal capabilities and limits of this approach.

After the preparation of the sample, large scale STM images of the Cu(110) surface exhibit large atomically flat terraces separated by single atomic steps. The terraces exhibit low amount of contamination. With such a clean surface, one would reasonably expect a homogenous distribution of LDOS across the terraces. However, similarly to the STS maps recorded on W(111) and Fe(001), the STS maps in that present case reveal a deviation of LDOS from its regular value in many locations. Figure 6 shows two STS maps of the same area scanned at two different biases: 100 mV (Fig. 6(a)) and 150 mV (Fig. 6(b)). The locations with deviating LDOS can be described as a system of spots of various sizes and shapes with different and variable contrasts. Some of the spots, the largest ones, are indexed with letters S1-S5 in figure 6 (a). Similarly to the W(111) and Fe(001) samples, presented above, the contrast of the spots in the same locations depends on bias voltage, can be low, for example the spots S1 and S5 at 100 mV (Fig. 6 (a)) or high (see the corresponding spots at 150 mV (Fig. 6 (b))), however this is also applicable to small spots. Differently from Fe and W samples, all

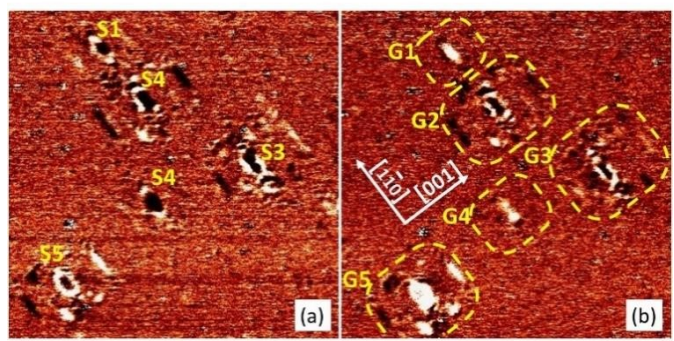

Fig. 6. Differential conductance distribution across the Cu surface (STS map) at two different bias voltages: (a) 100 mV and (b) 150 mV. Tunnel current: 1.5 nA. Size: 45 × 45 nm$^2$. Spots located at same positions change their contrast with bias voltage. The spots can be grouped in the uniform ensembles (encircled with dashed line as labelled G1 to G5 in (b)) composed of main (labelled as S1 to S5 in (a)) and satellite spots. Satellite spots around some main spots of lowintensity somewhere are not well resolved.

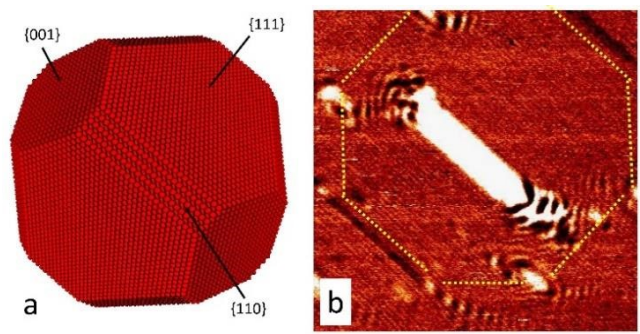

Fig.7 (a) Expected shape of the last layer of Cu interface around an Ar nanocluster; (b) High resolution STS map (16 x 16 nm$^2$) above an Ar nanocluster revealing the main oblong spot originated from the (110) facet, and the satellite spots corresponding to the conjunction of the {110}, {111}, and {001} facets, and the conjunction of the {111} and {001} facets. The yellow contour indicates the projected subsurface nanocluster.

the spots of different shape and size form reproducible groups encircled with dashed labelled with G1 to G5 in Fig. 6b. The groups are composed of a main oblong spot of the strongest contrast in the centre of the each group flanked by two elongated lines and surrounded by 4 more rounded small satellites spots (Fig. 6 and 7(b)). Remarkably, the oblong main spots as well as line-shaped satellite spots are all well oriented along [1-10] direction. If the main spots appear with a quite low contrast, for example the spots S1 and S4, the satellite spots may be vague at some bias voltages (Fig. 6 (a)) or absent at other bias voltages (Fig. 6 (b)). It would be natural to associate not each observed spot but each group of the main and satellite spots with a corresponding single Ar nanocluster hidden beneath the surface [35, 36]. The top view from the (110) surface of the idealized Ar nanocluster presented with the first interface Cu layer is depictured in figure 7(a). We demonstrated previously in our work [35] the correspondence between the observed spots and specific facets: the oblong main spot has been shown to be attributed to the parallel upper (110) facet that could reflect the injected electrons back. The satellite spots are induced by the ordered atomic structure in between the {110} and {001} facets whereas the satellite lines are associated with scattering at the edges of {111} facets [35]. In other words, each observed feature can be associated with QW resonances coming from specific facets of the nanocluster. The contour line connecting the small satellite spots and passing through satellite lines gives the exact size and shape of the whole nanocluster (Fig. 7(b)). As explained previously, the actual contour may appear slight asymmetrical due to deviations from the idealized shape of the nanocluster as in Fig. 7(a).

Since the origin of the spots is established, the measurement of oscillations of differential conductance only in the centre of main oblong spot associated with the closest (110) facet in each group is sufficient to determine the depth of Ar nanocluster. The sets of these measurements for different depths is presented in figure 8. These measurements are performed with the sample after the deposition of copper in increments as described above. Figure 8 shows different nanoclusters after each deposition step and selected to be of approximately equal size. By this process of fabrication and analysis we aimed at practical demonstration of ultimate depth of STM subsurface vision.

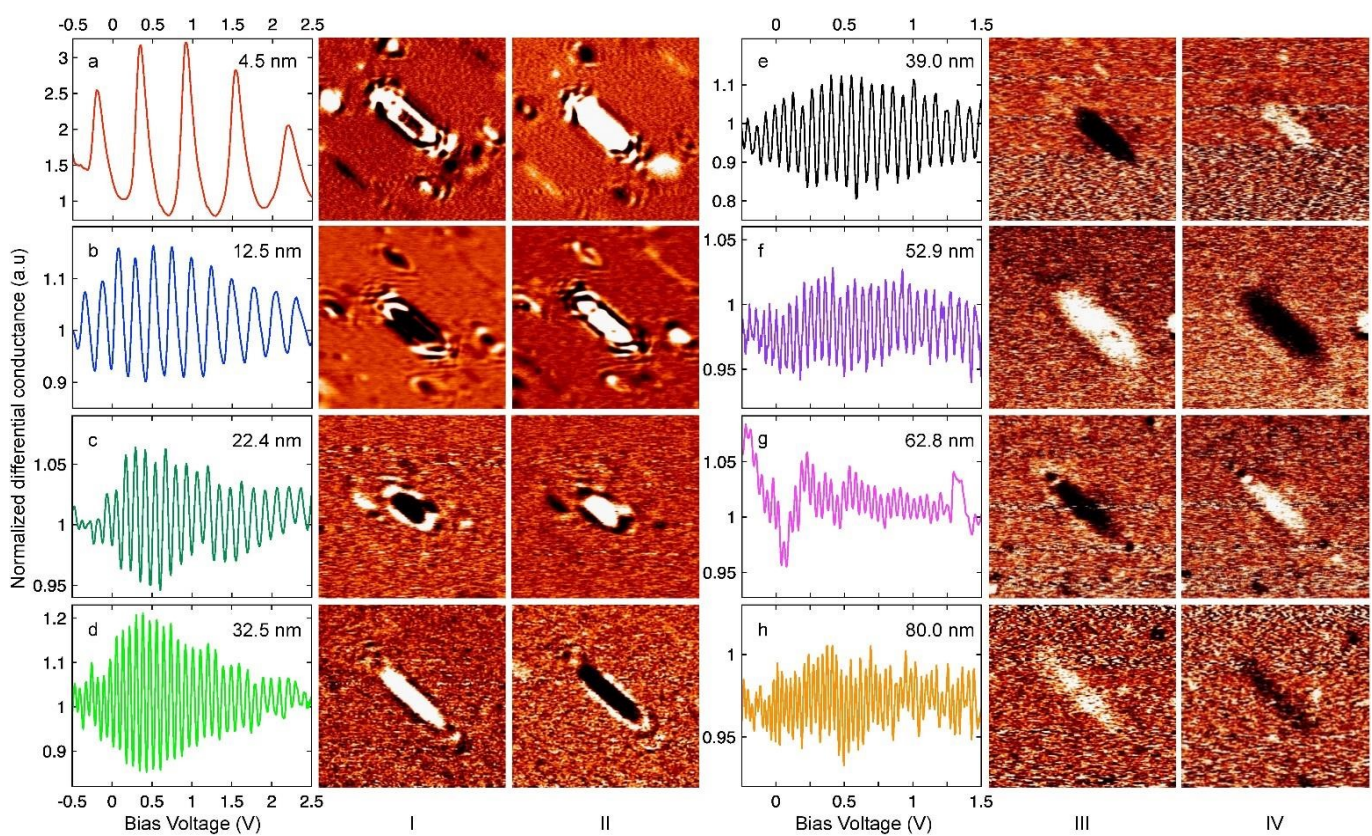

Fig. 8. Evolution of spatially resolved and normalized differential conductance as a function of burial depth: a) 4.5 nm, b) 12.5 nm, c) 22.4 nm, d) 32.5 nm, e) 39.0 nm, f) 52.9 nm, g) 62.8 nm, and h) 80.0 nm. For each specific depth (a - h), tree panels show: oscillating normalized differential conductance versus bias voltage measured in the centre of the main spot, and two maps of differential conductance (columns I, II (a-d) and columns III, IV (e-h)) recorded at two different bias voltages showing the corresponding spot with opposite contrasts. The contrast and oscillation amplitude at the depth of 80 nm (h) are high enough to detect the buried nanocluster. The shape and size of the main spot can be used to judge the size of the deep subsurface nanocluster if the satellite spots are not observed. Measurements a-d and e-h are performed at 77K and 4.5K, respectively and with different ranges of bias voltage.

**Nanoscale Horizons**

**ARTICLE**

First, the differential conductance measured in the centre of main spot for all the depths (Fig. 8, see plots) clearly reveal remarkable oscillations versus bias voltage with the periodicities specific for each depth. Similarly to Fe and W samples, this directly proves the connection of the observed spots with the subsurface scattering or QW resonances. Additionally, two STS maps with the corresponding buried nanocluster are shown on the right of each plot of oscillating normalized differential conductance in figure 8. The maps were recorded at two voltage biases corresponding to one maximum and one minimum of the differential conductance and therefore the main spots demonstrate the opposite contrasts. This also gives better understanding of the variety of contrasts for different spots observed in Fig. 6: due to the difference in the depths of location of nanocavities, the confined electronic states are in resonance or out of resonance at the chosen bias voltage. The period of oscillation determines the depth of location of the reflective facet calculated by formula 1 using tabulated data on $E(k)$ for copper [48]. The calculated depth is indicated in the upper right corner of each STS spectrum and ranges from 4.5 to 80 nm. Note that the STS spectra and maps in Fig.8 (a – d) were obtained at 77 K, whereas those presented in Fig. 8 (e – h) were measured at 4.7 K for deeper nano-objects. The calculated depth is consistent with the estimated one calculated by summing Ar implantation depth as described in the Experiment section which is within the typical range of 1 nm to 10 nm [49] and the resulting thickness of the copper layer added in increments of 10 nm. The satellite spots in figure 6 are well visible (a - c) at the corresponding depths of 4.5 nm, 12.5 nm and 22.4 nm, respectively, and still recognisable in the pattern (d) at the depth of 32.5 nm. The contour formed by these satellite spots and lines can reveal the size of the buried nanocluster (see similar indication in fig. 7). However, the satellite spots are no longer visible for the deeper nanoclusters, although the main spot is still observable for the depths down to 80 nm (Fig. 8 (e – h)). Their lack of visibility of the satellite spots makes impossible to estimate the size and shape of the nanoclusters buried deeper than 30 nm by the way we described above. However, the shape and size of the oblong main spot corresponding to the electron scattering from the upper (110) facet remains well visible and keeps the same aspect ratio even for depths down to 80 nm (Fig. 8 (h)). Actually, this is sufficient to estimate, albeit less accurately, the size of the hidden nanocluster using an alternative assumption without using the satellite spots. If we assume the constant ratio between the size of the {110} facets and the size of the other facets in the Wulff construction at the different depths, the size of the entire nanocluster can be estimated by only knowing the size of one facet. It is reasonable to assume, as the interface energy is not expected to depend on the location of the nanoclusters in the bulk. Usually, near-surface relaxation, which may affect the interface tension, decays within one nanometer [50] while all our Ar nanoclusters lie much deeper. This should support our assumption.

In any case, we have gotten a direct and impressive confirmation of subsurface STM vision really achieved for the depth of 80 nm (Fig. 8h). Since the signal-to-noise ratio is higher than 1 in both the oscillation plot and the pair of STS images, the STM is able to see even deeper.

In order to determine the ultimate depth of the STM subsurface vision, we have plotted in Fig. 9 the variation of the oscillations amplitude $Q$ of the normalized differential conductance as a function of depth $d$ recorded at the centre of the oblong spot. Indeed, the interpolation of such a plot would give an estimate of such ultimate depth. Data are presented in Fig. 9 for many nanoclusters of identical size to discard size dependency. The measurements were performed at two temperatures: at 77K for shallow nanoclusters and at 4.7K for deeper ones. As a consequence, the decay rate must be adjusted for these two temperatures separately. For clarity, data taken at 77K and 4.7K are indicated in blue and red, respectively in Fig. 9.

In Fig. 9, if one discard the two sets of data around a depth of 12 and 22 nm as will be discussed below, the intersection between the actual noise level horizontally drawn in green in Fig. 9 with the decay rate interpolation would lead to an ultimate depth of roughly 110 nm at 4.7 K and 70 nm at 77 K. The strong attenuation of $Q$ for these two abovementioned sets of data could be explained by different tip termination states that could strongly alter the sensitivity to the oscillatory component of the tunnelling signal. Unfortunately, a particular electronic state at the tip apex is not always controllable and can depend on tip or can be changed over time by intentional or accidental perturbations. Therefore, the ultimate depth deduced from the plot in Fig. 9 may be underestimated and a perfect STM tip with an appropriate electronic state could provide much better results.

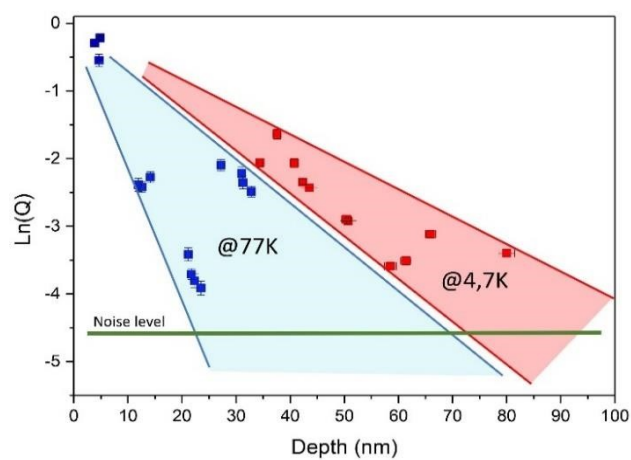

Fig.9 Amplitude of oscillatory contrast Q of the normalized conductance versus depth $d$ measured at 77K (in blue) and 4.7K (in red). Each point corresponds to different nanoclusters. Grouping the points with low oscillation amplitude in the range of 10 nm to 25 nm is due to the variation of the tip sensitivity to the oscillatory component of electronic states in some sets of measurements.

The two decay rates at two temperatures indicate that electron-phonon scattering processes influence the coherence of the electronic states. Another temperature effect usually considered when performing STS studies is the thermal broadening of the resonance states. However, in our case, it plays an insignificant role since the periodicity of STS oscillations in our experiments is always much higher than the value of $k_BT/e$ (with $k_B$, Boltzmann's constant).

The reported ultimate depth could be considered as a record achievement for some exceptional case. Indeed, the favourable combination of a long MFP, a specific band structure providing the focusing effect, and a total reflection from a smooth interface with the nanoclusters yields this record value for Cu(110) while other systems would provide a moderate ultimate depth. Among all the components of this combination, the role of MFP as the scaling parameter is clear, while the contribution of interface quality and focusing is not well-defined. To check the real influence of these two factors we used another system which have more practical importance for nanomagnetism and spintronics, namely the Co and Fe nanoclusters buried in Cu(001), Cu(110) and Cu(111). The system of Co nanoclusters in Cu(001) is reported in the next sub-section, whereas similar results on Fe nanoclusters in Cu(001) are shown in Supplementary Materials. The results obtained with other orientations of copper sample are briefly discussed at the end of the following sub-section as well.

## 4. Cu(001) with embedded Co nanoclusters: ultimate depth without focusing

STS mapping of the flat surface reveals many locations with deviating conductance (Fig. 10 (a-c)). Disregarding point defects originating from surface flaws and inclusions, relatively large spots about 5 nm are visible in the STS images. These spots are surrounded by interference rings, while the contrast of the spots depends on bias voltage. These spots are associated with the buried Co nanoclusters. The method of preparation of the sample by forming Co nanoclusters on the surface and their burying (see "Experiment" section) provides mostly the equal depth for all nanoclusters. Therefore, the spots show the same contrast within a chosen ensemble. However, somewhere some spots can exhibit different contrast (Fig. 10 (b)) thus implying a slightly different depth that can originate from to the stepped surface. The STS measurements recorded at the centre of a selected spot reveal oscillations in normalized surface conductance similarly to the systems considered above. The spots on the STS maps and the oscillations in LDOS have been registered even for deeper nanosclusters (not shown here). However at 25 nm only a very faint pattern can be observed (Fig. 10 (d)) with the corresponding oscillations of the surface conductance reduced to the noise level (Fig. 10 (h)). Therefore, we can deduce that the ultimate depth of subsurface vision for Co nanoclusters in Cu(001) matrix is 25 nm. The same ultimate depth is obtained with Fe nanoclusters in Cu(001) (see Supplementary Materials). The measurements have been carried out at 77K. Therefore, the ultimate depth of 25 nm determined for this non-ideal system should be compared with the ultimate depth of 70 nm for the more efficient system of Ar in Cu(110) kept at the same temperature. This comparison clearly shows that the partial reflection and the absence of focusing reduce the ultimate depth by approximately a factor 3.

Nevertheless, the ultimate depth of the subsurface vision of 25 nm for a non-ideal system is still highly impressive.

Our attempts to get the subsurface STM vision of Co nanoclusters buried in Cu(111) and Cu(110) crystals did not lead to acceptable results. The first case is consistent with other observations, even using the ideal Ar nanoclusters because the Cu(111) cut does not provide the corresponding electronic states in the <111> directions. As a result, the QW states, which we want to exploit for the subsurface vision, cannot be formed with this direction cut. On the contrary, in the case of the (110) orientation of Cu, the absence of QW for Co nanoclusters buried into Cu(110) has another explanation. Despite the focusing leading to a large depth of detection as in the case of Ar nanoclusters, QWs cannot be formed most probably due to a rough interface. Indeed, partial intermixing at the Cu(110) interface is more likely than on other orientations [51, 52] thus destroying the coherence at reflection and preventing the formation of a QW state. From these observations, one can conclude on the paramount importance of the coherence provided by a smooth interface, while the presence of focusing or partial reflection plays a secondary role in the realization of the subsurface STM vision.

## 5. Generalization of subsurface STM vision

The subsurface STM vision using near-surface QWs can be realised with a variety of materials as partially proven in this article and in the Supplementary Materials. The generalization of our approach could also be enlarged considering other combinations of materials such as Cd, Cu, Pb, Ag or Al where near-surface QW observations have been reported [16-24, 29 - 36]. Many other materials have yet to be tested.

The main condition to get the subsurface vision is the formation of near-surface QW states with *k*-vectors in the direction perpendicular to the surface and the interface. Despite some exceptions, the electronic states with perpendicular *k*-vectors are almost always available considering various metals in many crystalline directions. Therefore a general requirement in getting QW states is keeping coherence confined electronic states the direct and reflected electron waves. The coherence

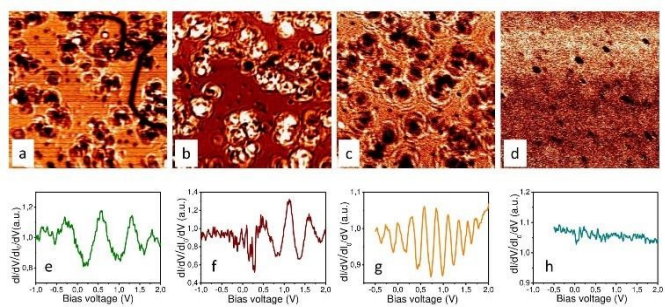

Fig. 10 (a – d) STS maps of Cu(001) sample with embedded Co nanoclusters buried at : (a) -3 nm, (b) – 4.8 nm, (c) – 10 nm, (d) – 25 nm. Size: 40 x 40 nm$^2$ ; (e –h) oscillatory STS plots measured above selected nanoclusters buried at the depth indicated in (a-d) for corresponding samples, respectively.

**Nanoscale Horizons**

**ARTICLE**

length of electronic states should be at least twice larger than the confinement length. The coherence length is generally limited by MFP ranging from a few nanometers to several tens of nanometers, that sets the ultimate limit of the subsurface STM vision. Since MFP depends on the type of metals, the concentration of impurities and the temperature, the selection of perfect crystalline materials and cooling down favours for the subsurface STM vision at bigger depths as we have demonstrated experimentally. The partial confinement by half-transparent interface is sufficient for formation of near-surface QW. This broadens the application including the use of conductive nanostructures.

However, the coherence of the reflected electronic waves could be destroyed at a rough interface. A rough interface can appear at some specific crystalline orientations or due to intermixing of the two materials. This also has been confirmed in our experiments.

The focusing, on the other hand, is not as crucial for achieving the subsurface STM vision, although this effect can increase the ultimate detection depth by the factor two or three and enhance the resolution. The have illustrated the influence of this effect experimentally.

Since the system represented by Ar nanoclusters in Cu(110) has the superior crucial parameters, analysing this system we can reasonably conclude that we have found the real practical limit for the subsurface STM vision. Nevertheless, even if the depth of subsurface characterization with other type of materials would be several times less than 100 nm, we can still claim that the QW-assisted subsurface STM vision is a promising method for many applications.

## Conclusions

We have presented the concept and the experimental proof of deep subsurface vision using STM/STS. Nano-objects hidden below the surface as deep as several nanometers up to 110 nm can be characterised by STM. In that sense, in spite of commonly use of STM as a surface sensitive method, this approach brings bulk sensitivity to this characterization tool. Our approach is based on the use of near-surface QW states formed in the host material due to the electron confinement between the surface and the buried nano-objects. These QW states probed on the surface by STS reveal periodic oscillations of the LDOS. The spatial variation of the oscillatory LDOS carries information about the location, size, and, in some cases, shape of the buried nano-objects while the oscillation period of the LDOS defined by the QW width, determines the depth at which the nano-obgects are buried. The effect of electron focusing which originates from a specific band structure of the host material enhances the ultimate depth of the STM subsurface vision. These results show that STM is well suitable for non-destructive 3D characterisation of hidden nano-objects present near the surface. Our approach can be applied to various combination of materials.

## Author Contributions

O. K. and M. S. contributed equally to this work. O.K. carried out the low-temperature-STM/STS experiments and treated the results. O.K. and M.S. initially wrote the manuscript and all the authors contributed. E.G. performed the theoretical calculations of band structure and Wulff constructions. O. K. did the sample fabrication at the assistance of D. P.; Y. L. fabricated Fe samples and helped at interpretation of QW results. S.M. and O.K. supervised the work. All authors approved the final version of the manuscript.

## Conflicts of interest

There are no conflicts to declare.

## Acknowledgements

This work was supported by the French PIA project "Lorraine Université d'Excellence" reference ANR-15-IDEX-04-LUE, and by the project CAP-MAT supported by the "FEDER-FSE Lorraine et Massif Vosges 2014–2020", a European Union Program. O.K. was supported by the region Grand Est under the DESOMIN project. O.K. and S.M. acknowledge the financial support of the Institut Carnot ICEEL. We are thankful to Can Avci and Bert Koopmans for their interest at preliminary results, and to Mattijn Cox for offering a graphical illustration for figure 1. We also want to recollect the memory of Karl-Heinz Rieder (1942 – 2017) who appreciated our very first results in private communications and inspired us for going further. Computing resources were provided by the EXPLOR centre hosted by the University de Lorraine.

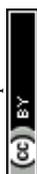